\PassOptionsToPackage{unicode}{hyperref}
\PassOptionsToPackage{hyphens}{url}
\documentclass[
]{amsart}
\usepackage[maxbibnames=99, backend=biber, firstinits=true]{biblatex}
\usepackage{hyperref}
\usepackage{geometry}
\usepackage{amsthm}
\usepackage{tikz}
\usetikzlibrary{positioning, calc, backgrounds, fit, shapes.misc}

\usepackage[ruled]{algorithm2e}

\theoremstyle{definition}
\newtheorem{prop}{Proposition}%[section]

\newtheorem*{lem*}{Lemma}

\newtheorem{dfn}[prop]{Definition}

\begin{document}
\author[W.~Atherton]{William Atherton}
\address{Department of Computer Science, University of Oxford}
\email{william.atherton@keble.ox.ac.uk}

\author[D.~Pasechnik]{Dmitrii V. Pasechnik}
\address{Department of Computer Science, University of Oxford} 
\email {dima.pasechnik@cs.ox.ac.uk}

\date{}
\title{Decline and Fall of the ICALP 2008 Modular Decomposition algorithm}

\begin{abstract}
We provide a counterexample to a crucial lemma in ICALP 2008 paper
\cite{10.1007/978-3-540-70575-8_52}, invalidating the algorithm
described there.
\end{abstract}

\maketitle

\section{Introduction}
Graph modular decomoposition is an important technique in graph theory,
and a well-studied algorithmic problem,
with dozens of different algorithms published since the pioneering
work \cite{Gal67}, see e.g. \cite{HABIB201041}. One highly cited
linear time algorithm has been presented at ICALP 2008 \cite{10.1007/978-3-540-70575-8_52},
also publised as \cite{tedder2008}. 
A Java implementation \cite{java2008} has been made available by the first author of [loc.cit.].
An archival copy of this implementation can be found in \cite{teddercode}.
As well, a number of attempts has been made to implement it in a different language,
one of them by the first author of this note, as a part of an undergraduate project \cite{Will2023}
supervised by the second author in Spring of 2023. At the testing phase of the implementation, done in SageMath \cite{sage}, an example,
see Sect.~\ref{sec:example},
has been found which produced an obviously incorrect output. The problem
was traced back to a lemma in \cite{tedder2008}, which is invalidated by the example.
As well, the implementation \cite{java2008} run on this example produced
basically the same incorrect output.

We contacted the authors of \cite{tedder2008} in May 2023, who were quick to
acknowledge the problem to us. In March 2024 they published a revision \cite{corneil2024recursive} of
\cite{tedder2008}, without mentioning the problem in [loc.cit.] and in the associated with it implementation
\cite{java2008}, and
without mentioning our communication.
The main purpose of this note is to publicise the problem in \cite{tedder2008}, \cite{10.1007/978-3-540-70575-8_52},
and thus to stop further waste of time stemming from attempts to implement the incorrect algorithm.

\tikzset{node distance=20mm,on grid,
         vert/.style={circle,draw=black!50,minimum size=7mm,
	 thick}}
       
\newcommand{\blueww}{\tikz[baseline=(bw.base)]\node[draw=blue](bw) {blue};}
\newcommand{\redww}{\tikz[baseline=(rw.base)]\node[draw=red](rw) {red};}
\newcommand{\greenww}{\tikz[baseline=(X.base)]\node[draw=green](X) {green};}

\begin{figure}\label{fig:G}
\begin{tikzpicture}[scale=.7, transform shape]

\node[vert] (a) {$a$};
\node[vert]  (i) [below of=a] {$i$};
\node[vert]  (b) [right=55mm of a] {$b$};
\node[vert]  (e) [below=1cm of b] {$e$};
\node       (ec) [below of=b] {}; % empty node
\node[vert]  (c) [below of=ec] {$c$};
\node[vert]  (d) [below left of=a] {$d$};
\node[vert]  (f) [below of=i] {$f$};
\node[vert]  (g) [left of=ec] {$g$};
\node[vert]  (h) [right of=ec] {$h$};

\foreach \x in {f,a}
   \foreach \y in {b,g,c,h}
      {\path [draw] (\x)--(\y);}
\path [draw] (a)--(e);
\path [draw] (f) to [bend right=20] (e);
%deal with (i)
\foreach \y in {b,g,c,e}
   {\path [draw] (i)--(\y);}
\path [draw] (i) to [bend right=10] (h);

%deal with (d)
\foreach \y in {b,g}
   {\path [draw] (d)--(\y);}
\path [draw] (d) to [bend right=30] (c);
\path [draw] (d) to [bend left=0] (e);
\path [draw] (d) to [bend left=10] (h);

\path [draw, very thick] (a)--(i)--(d)--(a);
\path [draw, very thick] (e)--(b)--(g)--(c)--(h)--(b);

\node[draw=blue,   fit=(a) (i) (d)] (3clique) {};
\node[draw=blue] [left=10mm of a] {series};
%\node[below, draw=blue] at (3clique.south) {series};

\node[draw=red,   fit=(3clique) (f)] (3plus1) {};
\node[draw=red] [left=10mm of f] {//};

\node[draw=red,   fit=(g) (h)] (gh) {};
\node[draw=red] [below of=b] {//};

\node[draw=green,   fit=(gh) (b) (c) (e)] (C4plus1) {};
\node[draw=green] [below of=h] {prime};

\node[draw=blue, fit=(C4plus1) (3plus1)] (rootG) {};
\node[below, draw=blue] at (rootG.south) {series};

\end{tikzpicture}
\caption{The modular decomposition of $G$, with "series",
resp. "parallel" (abbreviated "//"), resp. "prime",
nodes of the decomposition tree are show by
\protect\blueww, resp. \protect\redww, resp. \protect\greenww\ boxes.
}
\end{figure}
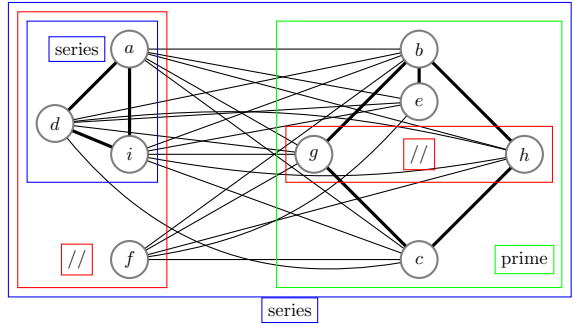

We quickly recall that for a graph $G$ with the vertex set $V:=V(G)$, a \emph{module} is a subset of vertices $M\subseteq V$
s.t. $M$ any two $x,y\in M$ cannot be distinguished by any $v\in V\setminus M$, i.e.
$v$ is either simultaneously adjacent to $x$ and $y$, or non-adjacent to $x$ and $y$.
Modules $V$ and singleton modules are called \emph{trivial}.
Note that connected components of $G$ and of the complement graph $\overline{G}$ are modules.
A module $M$ can be properly contained in another module $M'\neq M$, or they can \emph{overlap}, in the sense that
$M\cap M'\neq\emptyset$, but neither $M'\subset M$, nor $M\subset M'$. A module which does not overlap any module is
called \emph{strong}. A \emph{modular decomposition tree} is a recursive partition (also known as a laminar set family) of $V$
into strong modules; an example may be see on Fig.~\ref{fig:G}. More details may be found in e.g. \cite{HABIB201041}.

\section{The example}\label{sec:example}
The counterexample graph $G$ which was discovered is described here. Note that not only its isomorphism type matters
here, but the particular vertex ordering---different orderings lead to different outcomes, for some of them
the result is correct. We mark nodes with letters $a$, $b$,\dots, $i$, to adhere to the vertex labeling conventions
of \cite{java2008}.
$G$ is the join of two graphs: the 4-cycle with an extra edge attached at a vertex, and the disjoint union of a vertex and
the 3-cycle. $G$ and its modular decomposition is shown on Fig.~\ref{fig:G}.

The implementation \cite{java2008} on this example produces
{\footnotesize
\begin{verbatim}
(SERIES, numChildren=5
    (label=e, neighbours:a,b,d,f,i), (label=b, neighbours:a,d,e,f,g,h,i),
    (label=c, neighbours:a,d,f,g,h,i),
    (PARALLEL, numChildren=2
        (label=g, neighbours:a,b,c,d,f,i), (label=h, neighbours:a,b,c,d,f,i)),
    (PARALLEL, numChildren=2
        (label=f, neighbours:b,c,e,g,h),
        (SERIES, numChildren=3
            (label=j, neighbours:a,b,c,d,e,g,h), (label=a, neighbours:b,c,d,e,g,h,i),
            (label=d, neighbours:a,b,c,e,g,h,i))))
\end{verbatim}
}
\noindent
which is obviously incorrect, there is no prime node!
Or, alternatively, note that the top series node
contains 5 children, i.e. the corresponding quotient graph is $K_5$,
and in particular there must be an edge between the singleton
nodes $b$ and $c$. The latter would only be possible if
$(b,c)$ was an egde in $G$---which is not the case.

Alternatively one could compute the decomposition of the complement
$\overline{G}$ of $G$. Such decompositions must be ``dual'' to each other,
in the sense that one must swap meanings of ``series'' and ``parallel``.
Indeed, \cite{java2008} works correctly on $\overline{G}$ and
produces, modulo the swap just mentioned, the decomposition at
Fig.~\ref{fig:G}.
{\footnotesize
\begin{verbatim}
(PARALLEL, numChildren=2
    (SERIES, numChildren=2
        (label=f, neighbours:a,d,i),
        (PARALLEL, numChildren=3
            (label=i, neighbours:f), (label=a, neighbours:f), (label=d, neighbours:f))),
    (PRIME, numChildren=4
        (label=b, neighbours:c), (label=c, neighbours:b,e), (label=e, neighbours:c,g,h), 
        (SERIES, numChildren=2
            (label=h, neighbours:e,g), (label=g, neighbours:e,h))))
\end{verbatim}
}

\section{Faulty lemma}\label{sec:faultylemma}
We were able to trace the flaw down to Lemma 4 in
\cite{10.1007/978-3-540-70575-8_52}
(which is Lemma 3.1 in the preprint version \cite{tedder2008}). The latter
lemma plays a crucial role in the proof of correctness and is shown
false on our example.

Let $x$ be an arbitrary vertex of $G$.
\begin{lem*} [Lemma 4 in \cite{10.1007/978-3-540-70575-8_52}]
The nodes in the ordered list of
trees resulting from refinement that do not have marked children
correspond exactly to the strong modules {\bf not} containing $x$.
\end{lem*}
First, note a typo in the statement of Lemma 4, which we corrected
above in {\bf boldface} - the missing ``not''.

This is a misprint. The proof of the Lemma is clearly
proving the statement with ``not'' inserted. As well,
the way it is used in the proof of Lemma 3 in
\cite{10.1007/978-3-540-70575-8_52} also indicated the correct
statement should have the missing ``not''.

We now show where the algorithm is going wrong on $G$.
It starts by choosing a vertex $x$ to be the first {\em pivot} (this
choice is arbitrary in the algorithm), and recurses on $G(x)$,
the set of the neigbours of $x$, a tentative (strong) module.
Then it processes the non-neighbours $\overline{G(x)}$,
resulting in a number of tentative
modules as in \eqref{eq:treeList}, and, finally, does a refinement step:
rearranging tentative modules into the modules for the tree. 
For a detailed complete description, see \cite{10.1007/978-3-540-70575-8_52}.

\begin{equation}  \label{eq:treeList}
\underbrace{T(N_0)}_{G(x)},x,\underbrace{T(N_1),\ldots,T(N_k)}_{\overline{G(x)}},
\end{equation}

Let the algorithm choose the vertex $i$ to be the first {\em pivot}.
We get
the neighbours of $i$, $G(i)=\{a,b,c,d,e,g,h\},$
$\{i\}$, and $\overline{G(i)}=\{f\}$.

It then recursively processes the neighbour partition.
The recursively computed modular decomposition for $G(i)$
can be seen on Fig.~\ref{fig:G}; one has to remove $i$, $f$, and the
two nontrivial decomposition tree nodes (strong modules) containing $i$
(i.e. the two nested boxes on the left of the green box).

The next step of the algorithm, ``pull-forward'', is skipped,  as
there is only one node, $f$, in $\overline{G(i)}$.

It then calculates the modular decomposition of $\overline{G(i)}$,
which is just the single-vertex tree consisting of $f$.

It then goes on to the refinement step, where the error lies.

The refinement process consists of Algorithms 1 and 2
from \cite{10.1007/978-3-540-70575-8_52}, which we copy here verbatim.
We process the following decomposition of the vertices of $G$ into
subtrees of modules.
\begin{algorithm}

\ForEach{vertex $v$} {

        Let $\alpha(v)$ be its incident active edges\;
        
        Refine the list of trees using $\alpha(v)$ according to algorithm~\ref{alg:2}, such that:
        
        \uIf{$v$ is to $x$'s left} {
                refine using left splits, and when a node is marked, mark it with ``left''\;
        }
        \uElseIf {$v$ is to $x$'s right and refines a tree to $x$'s left} {
                refine using left splits, and when a node is marked, mark it with ``left''\;
        }
        \ElseIf {$v$ is to $x$'s right and refines a tree to $x$'s right} {
                refine using right splits, and when a node is marked, mark it with ``right''\;
        }
}
        
\caption{Refinement of the ordered list of trees~(\ref{eq:treeList}) by the active edges} \label{alg:1}
\end{algorithm}

\begin{algorithm}

Let $T_1,\ldots,T_k$  be the maximal subtrees in the forest whose leaves are all in $X$\;

Let $P_1,\ldots,P_\ell$ be the set of parents of the $T_i$'s\;

\ForEach{non-prime $P_m$} {

        Let $A$ be the set of $P_m$'s children amongst the $T_j$'s, and $B$ its remaining children\;

        Let $T_a$ either be the single tree in $A$ or the tree formed by unifying the trees in $A$ under a common root, and define $T_b$ symmetrically\;
        %Create a new tree $T_a$ formed by unifying the trees in $A$ under a common root\; 
        %Create a new tree $T_b$ formed by unifying the trees in $B$ under a common root\; 

        Assign $P_m$'s label to $T_a$ and $T_b$\;

        \uIf{$P_m$ is a root} {
                Replace $P_m$ in the forest with either $T_a,T_b$ (left split) or $T_b,T_a$ (right split)
                %\tcc*{specify left or right when invoking the algorithm}
        }
        \Else {
                Replace the children of $P_m$ with $T_a$ and $T_b$\;
        }
        
        %If either $T_a$ or $T_b$ has a single child, delete the root of the tree and make its child the new root\;
        Mark the roots of $T_a$ and $T_b$ as well as all their ancestors\;
}

\ForEach{prime $P_m$} {
        Mark $P_m$ as well as all of its children and all of its ancestors\;
}

\caption{Refinement of an ordered list of trees by the set $X$} \label{alg:2}
\end{algorithm}

Lemma 4 is necessary for the correctness of the main
algorithm. However, we will be
going through Algorithm \ref{alg:1} on the given example $G$, and show
that its result contradicts Lemma 4.
We start its loop from vertex $f$.
Then $v=f$ and
$\alpha(f)$ is the list of incident active edges of $f$.
\begin{dfn}[\cite{10.1007/978-3-540-70575-8_52}]
An edge becomes \emph{active} when one of its endpoints is a
pivot or if its endpoints reside in different layers.
\end{dfn}

As $f$ is in its own layer, all its incident edges are active.
Therefore $\alpha(f) = \{bf,cf,ef,gf,hf\}$.
Algorithm \ref{alg:2} is then run on the ordered list of trees, with the set
$X = \{b,c,e,g,h\}$.

It first calculates
$T_{1},\ldots,T_{k}$, the maximal
subtrees in the forest whose leaves are all in $X$.
The list of maximal subtrees in this case is a single subtree
corresponding to the whole of $X$ is given by 
the prime node in Fig.~\ref{fig:G}, which we refer to as
$T_{1}$.

$P_{1}$, the
parent of $T_{1}$, is the topmost node from the modular
decomposition of $G(i)$.

As there is only one $P_{k}$, the outer loop is only
run once, on $P_{1}$.

$A$ is the set of $P_{1}$'s
children among the $T_j$'s, which is just the
singleton set $T_{1}$, and $B$ is the set of
remaining children, which is the singleton---the bottommost
parallel node from the modular decomposition of $G(i)$.

As $|A|=|B|=1$, we have that $T_{a}=A=T_1$, and $T_{b}=B$.

$P_m$ is a root, so $P_m$ is
replaced by $T_{a}, T_b$, as the
subtree is to the left of the pivot.

Then, the roots of $T_a$ and $T_b$
are marked, and so are all their ancestors.

This means that the root of $T_a$, the prime node at
the top of $T_1$, is marked.

The algorithm then marks the children of all prime nodes marked this
way, so the children of the prime node at the top of
$T_1$ is marked, meaning the nodes $b,$ $c,$ $e,$ and the
parallel node from $T_1$ are all marked.

This is a contradiction to Lemma 4. Lemma 4 states that the nodes in the
ordered list of trees resulting from refinement that do not have marked
children correspond exactly to the strong modules not containing $x$. This
means that the strong modules not containing $i$ must not have a marked
child. However, as you can see from Fig.~\ref{fig:G}, 
$T_1$ is a strong module not containing $i$, but it has
marked children. \qed

%Lemma 4 is used in the proof of Lemma 3, which is then used in the
%correctness proof.

\section{Conclusion}
From the implmentation point of view, the fact that the children of
$T_1$ get marked means that when the Promotion step
happens, the children of $T_1$ get split from
$T_1$, and the $T_1$ node gets
deleted. As the rest of the algorithm assumes that strong modules not
containing $x$ are not affected by refinement, these nodes do not get
reassembled back into a prime node, so you get the error occuring in the
example implementation, where the prime node is missing.

This is a fundamental problem with the algorithm, as Lemma 4 is used to
prove correctness of the algorithm, and the fact that children of prime
nodes get marked in Lemma 2 is important for other cases of the
algorithm to work correctly. Apparently the idea is not easy to salvage,
as \cite{corneil2024recursive} appears to take a quite different approach,
using LexBFS.

\printbibliography[heading=bibintoc]
\end{document}